\documentstyle[aps,eqsecnum,prbbib]{revtex}
\begin{document}
\draft
\title{Low Frequency Nonlinear Magnetic Response of an Unconventional 
Superconductor}
\author{Igor \v{Z}uti\'c\cite{igor}  and Oriol T. Valls\cite{oriol}}
\address{School of Physics and Astronomy and Minnesota Supercomputer Institute
\\ University of Minnesota, 
Minneapolis, Minnesota 55455-0149}
\maketitle
\begin{abstract}
We consider an unconventional superconductor in a low frequency harmonic 
magnetic field. In the Meissner regime at low temperatures a nonlinear 
magnetic response arises from quasiparticle excitations near 
minima in the energy gap.  As a consequence various 
physical quantities acquire higher harmonics of the
frequency of the applied ac field. We discuss how 
examination of the field and angular dependences of these harmonics allows 
determination of the structure of the energy gap. We show how to distinguish 
nodes from small finite minima  (``quasinodes''). Gaps with nodal lines, 
give rise to universal power law field dependences for the 
nonlinear magnetic 
moment and the nonlinear magnetic torque. They both have 
separable temporal and angular dependences. In contrast, with 
gap functions which only have quasinodes,
these physical quantities do not 
display power laws in the applied field,  and their temporal and 
angular dependences are no longer separable.
We illustrate  this via the example of  the nonlinear 
magnetic moment for a $d+is$  gap. We 
discuss how to 
perform  ac measurements so as to maximize the nonlinear signal,
and how to 
investigate in detail the properties of the superconducting minima, thus 
determining the gap function symmetry. 
\end{abstract}

\section{Introduction}
 
There are strong indications from numerous experimental results and 
theoretical calculations that the symmetry of the pairing 
state\cite{agl,dj,kirt,harl} in 
various superconducting materials is not that of an isotropic $s$-wave. 
Unconventional pairing states have been assigned to different High 
Temperature Superconductors (HTSC's), Heavy Fermions\cite{kum,upt} (HF) 
and some organic charge transfer salts\cite{org1,org2,org3} (OS).
Interest in determining the pairing state for these 
materials arises both from the efforts to obtain significant clues about the 
microscopic mechanisms responsible for superconductivity and to better
understand their physical properties. 

For HTSC's it is widely accepted that most of the experimental results 
support a predominantly $d$-wave symmetry. There is  however no
consensus about the presence of admixtures of pairing states of 
different symmetry which would modify the position and value
of the minima in the superconducting energy gap. These admixtures might
cause the angle  between the nodal lines to depart from $\pi/2$, or convert
the nodes to very deep minima (``quasinodes''), or both.
Part of the difficulties result from the 
surface character of many high quality pairing state probes. They measure 
information about the order parameter (OP) within a length scale of a few 
coherence lengths and are very susceptible to material imperfections 
near the surface.  Furthermore it is not 
clear whether the pairing states are the same in
the surface region as throughout the 
bulk.\cite{mul,bahc,bound,green,kouz} 
There are also indications that the symmetry of the OP 
might be temperature dependent.\cite{chu,ting,lambda}
The pairing state controversy  for other suggested
unconventional superconductors such as HF, ${\rm SrRu_2O_4}$\cite{rice,ish} 
and some OS is even more ambiguous.

In this paper we consider the low frequency magnetic response of a 
spin singlet 
unconventional superconductor in a {\it time dependent} magnetic field.
We focus on the low temperature  regime  in the Meissner 
state for OP's having lines of nodes
(or quasinodes.)
The nonlinear response\cite{ysprl,sv,ys,zv,jcp,zv3,lee,beto,aff} arising from 
quasiparticle excitations near the 
minima in the energy gap generates in various physical 
quantities higher harmonics\cite{scal} of the 
applied field frequency. Since the response
extends over a length scale on the order of penetration 
depth, it constitutes a bulk probe of the superconducting OP.  
The use of the nonlinear response to a time-independent field
 to perform gap spectroscopy
(that is, to locate the nodes or quasinodes in the gap) was 
previously discussed\cite{zv3}. However, 
nonlinear effects are best detected through the use of ac techniques,
since then the nonlinear response takes place at frequencies different
from the input frequency at which the much larger linear response is found.
These techniques significantly simplify the process of resolving the 
desired small nonlinear signal, 
which is a signature of the symmetry of the energy gap, from the large 
spurious but linear effects such as demagnetization factors, $a-b$ plane 
penetration depth anisotropy, and trapped flux. 

In Section II we solve the nonlinear Maxwell-London equations 
in the low frequency Meissner regime. We
generalize the perturbation method of Ref. \onlinecite{zv3} to 
include the temporal dependence. The method 
is illustrated on the example of OP with mixed $d+s$ symmetry. We 
investigate the time and angular dependence of the nonlinear magnetic 
moment and the associated torque. The results for these quantities are 
easily extended to other forms of energy gaps with nodes since, as
we shall see,  in those cases one has
separable temporal and angular dependences. The time dependence for gaps 
with lines of nodes is universal: $H_a(t)\left|H_a(t)\right|$ for the nonlinear 
magnetic moment and $\left|H_a(t)\right|^3$ for the nonlinear magnetic torque, 
where  $H_a(t)$ is the applied magnetic field. 
Both of these quantities have the 
same angular dependence. 
The nonlinear effects that we discuss here can be also viewed as field 
and angle 
dependent corrections to the superfluid density (penetration depth). 
We briefly discuss how
our methods are suitable to extend studies of intermodulation and 
harmonic generation\cite{scal} to low temperatures.

In Section III we consider superconducting gaps without nodes but
with quasinodes, as illustrated by a $d+is$ OP with a small $s$
component. We 
examine the nonlinear magnetic moment, which exhibits a more complicated 
temporal and angular dependence and a sharp enhancement of 
its maximum amplitude, compared to that occurring in the case of gaps with nodes. 
We show how to use these effects to
experimentally distinguish nodes from small minima in 
the superconducting gap.
In the final Section we present our conclusions and discuss possible 
extensions of this work.

\section{Nonlinear Magnetic Response}
\label{harm}
\subsection{Maxwell-London Electrodynamics}
In the low frequency regime, i.e. in the quasi-static 
case,\cite{scal,london,orlando} the relevant 
equations of the nonlinear Maxwell-London
electrodynamics are formally the same as in the static case.
Following the notation and results of  
the static or dc case,\cite{zv3} we have
\begin{equation}
\nabla\times\nabla\times{\bf v}=\frac{4 \pi e}{c^2}{\bf j(v)},
\label{maxlon}
\end{equation}
where the gauge invariant condensate
flow field or superfluid ``velocity'' ${\bf v}$ is defined as:
\begin{equation}
{\bf v}=\frac{\nabla \chi}{2} + \frac{e}{c} {\bf A},
\label{vdef}
\end{equation}
with  $\chi$ the phase of the superconducting singlet OP, ${\bf
A}$ the vector potential, and $e$ the proton charge.
The relation between ${\bf j}$ and ${\bf v}$ is generally
 nonlinear and given by\cite{ys,zv,zv3}
\begin{equation}
{\bf j(v)}=-eN_f\int_{FS} d^2s \: n(s) {\bf v}_f [({\bf v}_f 
\cdot {\bf v})
  +2\int^{\infty}_0 d\xi \: f(E(\xi)+{\bf v}_f\cdot {\bf v})],
\label{nonlinjv}
\end{equation}
where  $N_f$ is the total density of states at the Fermi level,
$n(s)$ the density of states at point $s$ at the Fermi surface (FS),
normalized to unity,
${\bf v}_f(s)$ the $s$-dependent Fermi velocity, $f$ the Fermi 
function, with $E(\xi)=(\xi^2+\left| \Delta(s) \right|^2)^{1/2}$,
$T$ the absolute temperature and $\Delta(s)$ the OP. The first term in 
Eq. (\ref{nonlinjv}) represents the supercurrent arising from the unperturbed 
condensate and the second is due to quasiparticle excitations.
At $T\approx 0$, and for lines of nodes
(or quasinodes), the second term of Eq. (\ref{nonlinjv}) can be written as\cite{zv3,fsa}
\begin{equation}
{\bf j}_{qp}{\bf (v)}=\sum_n {\bf j}_{qp,n}{\bf (v)}
\approx-2e \sum_n N_{fn}\int_{\Omega_n} \frac{d \phi_n}{2 \pi}
{\bf v}_{fn}
[({\bf v}_{fn} \cdot {\bf v})^2-\left| \Delta(\phi_n) \right|^2]^{1/2},
\qquad n=1,2,...
\label{jq}
\end{equation}
where $n$ labels the nodes (quasinodes) of the gap, $\Omega_n$ denote 
regions where quasiparticle excitations are allowed, defined by  
$\left| \Delta(\phi_n) \right|   +{\bf v}_f \cdot {\bf v} <0$,
$\phi_n$ is the azimuthal angle with respect to the closest node, $n$, and  
$v_f \approx v_{fn}$, its value at that node. $N_{fn}$ is the appropriate 
weighted density of states, \cite{weight} equal to $N_f$ for an 
isotropic FS. 

As stated in Ref. \onlinecite{scal}, the relation between 
${\bf  j}$ and ${\bf v}$ given by Eq. (\ref{nonlinjv}) can be used 
in the time dependent case provided that the  
frequencies are smaller than the quasiparticle relaxation rate. For 
${\rm YBa_2Cu_3O_{7-\delta}}$ (YBCO), this  rate ranges\cite
{bon,hir} from $10^{11}$ Hz to $10^{13}$ Hz, depending on the temperature. 
Measurements of the low frequency nonlinear magnetic response are 
designed to be performed\cite{bag} at frequencies 
under 100 Hz, well below this limit. 

For different OP's, we study the dependence of the nonlinear
magnetic response 
on the direction of the field applied in the $a-b$
plane. We consider a slab sample, infinite 
in the $a-b$ plane and of thickness 
$d$, much
larger than the in-plane penetration depth,\cite{thick}
in the $c$ direction. 
This allows for an analytic solution and 
preserves  the angular dependence of the nonlinear effects found for
realistic finite three dimensional geometry.\cite{zv,jcp}
For a slab, Eq. (\ref{maxlon})
can be written as
\begin{equation}
\partial_{zz}{\bf v}+\frac{4 \pi e}{c^2}{\bf j(v)}=0,
\label{lap}
\end{equation}
and the boundary conditions are:
\begin{equation}
{\bf H}(z,t)={\bf H}_a(t)|_{z=\pm\frac{d}{2}}.
\label{hatd}
\end{equation}
We take a particular time dependence of the applied field:
\begin{equation}
H_a(t)=H_{dc}+H_{ac} \cos \omega t,
\label{ht}            
\end{equation}
which is suitable to experimentally study higher harmonics arising from the 
nonlinear response. 
The parity of 
quantities such as $j_{qp}$ under a
sign change of $H_a(t)$, as it occurs in an ac field,
must be carefully taken into account.  For an OP satisfying 
$\left|\Delta(-{\bf k}_f)\right|=\left|\Delta({\bf k}_f)\right|$, where 
${\bf k}_f$ is the Fermi wave vector, and 
for ${\bf H}_a$ chosen such that the nodal direction of quasiparticle 
excitations is $\hat{x}_n$, the key point is that
a reversed field $-{\bf H}_a$ produces
excitations along the opposite direction, $-\hat{x}_n$. Therefore
$j_{qp}$ is odd. This will help us to 
anticipate various modifications of the results for the 
static case.

\subsection{Gaps with nodes}

We consider an energy gap which can 
be accurately approximated near its nodes\cite{ys,zv3} by 
\begin{equation}
\left| \Delta(\phi_n) \right|
\approx \left|\mu \Delta_{eff} \phi_n \right|, \qquad n=1,2,.. 
\label{gap}
\end{equation}
where $\mu$ is the slope of the 
dominant OP component and  $\Delta_{eff}$ is the effective amplitude of 
the gap function. A particular case of Eq. (\ref{gap}) corresponds to a 
$d \pm s$ OP of the form 
$\Delta_{d \pm s}(\phi)=\pm\Delta_s+\Delta_d\sin 2\phi$, 
where $\phi$ is measured from the $X$ axis, depicted in Fig. \ref{axes}.
This form is often chosen to include the effects of orthorhombic distortion in 
the YBCO family of cuprates, 
with $\mu=2$ and $\Delta_{eff}=(\Delta_d^2-\Delta_s^2)^{1/2}$.
For $\Delta_s \neq 0$, the nodes of Eq. (\ref{gap}) are no longer 
separated by an angle of $\pi/2$. 
The nodal directions are along the unit vectors $\hat{x}_n$, $n=1,2,..$ 
which form nonorthogonal axes. 
The nodes are shifted by an angle $\pm \nu$ (see
Fig. \ref{axes}),
\begin{equation}
\nu\equiv \frac{1}{2} \sin^{-1} (\frac{\pm \Delta_s}{\Delta_d}),
\label{nu}
\end{equation}
 from the orthogonal axes, which we denote by $X$ and $Y$ (the $\pm$ sign 
corresponds to the $d\pm s$ forms of the gap).
Throughout this paper the $X$ and $Y$ axes will remain along the nodal 
directions of a pure $d$-wave.
The direction of the applied field
 will be given by the angle $\psi$ between ${\bf H}_a(t=0)$ 
and the $+Y$ axis, depicted in  Fig. \ref{axes}. It is convenient to 
introduce dimensionless versions of the superfluid velocity and magnetic 
field. For simplicity, we will perform our
calculations below for an isotropic FS.
Our results are largely independent of this assumption, as
we will occasionally emphasize. It is however
straightforward to extend these considerations
and examine in detail the effects of FS
and $a-b$ plane penetration depth anisotropy as
have been addressed in Ref. 
\onlinecite{zv3}. We define:
\begin{mathletters}
\label{units}
\begin{equation}
u_i\equiv\frac{v_i v_f}{\mu\Delta_{eff}}, \qquad i=X,Y
\end{equation}
\begin{equation}
h_{dc, ac}=\frac{H_{dc, ac}}{H_0}, 
\quad H_0=\frac{c \Delta_{eff}}{e \lambda_{ab} v_f},
\quad h=h_{dc}+h_{ac}
\end{equation}
\end{mathletters}
where the in plane penetration depth, $\lambda_{ab}$, for a cylindrical 
FS is given by $2 \pi N_f v_f^2/c^2$.
In terms of these quantities and of the dimensionless coordinate 
$Z\equiv z/\lambda_{ab}$, we obtain, from Eqs. 
(\ref{jq}), (\ref{lap}) and (\ref{units}):
\begin{equation}
\partial_{ZZ}u_i-u_i
+{\rm sgn}(h(t))[e_{1i}(\psi) u_{X}^2+ e_{2i}(\psi) u_{X} u_{Y}
+e_{3i}(\psi) u_{Y}^2]=0, \qquad i=X,Y.
\label{ueq}
\end{equation}
where the factor ${\rm sgn}(h(t))$ arises from the 
odd parity of $j_{qp}$ with respect to $h(t)=h_{dc}+h_{ac}\cos \omega t$ 
and the constants $e_{1,2,3\;i}(\psi)$ are defined in Appendix \ref{ap1}. 
The appropriate boundary conditions for Eq. (\ref{ueq}) are
\begin{equation}          
\label{bcds}
\partial_Z u_{X}|_{Z=Z_s}=\frac{h(t)}{\mu}\cos \psi, \qquad
\partial_Z u_{Y}|_{Z=Z_s}=\frac{h(t)}{\mu} \sin \psi,  \qquad
u_{X,Y}|_{Z=0} \equiv 0,
\end{equation}
where $Z_s\equiv d/2\lambda_{ab}$.
The solution to Eq. (\ref{ueq}) and (\ref{bcds}) can be sought in the form:
\begin{equation}
u_i(Z,t)=u_{i0}(Z)+\sum_{j}[u^e_{ij}(Z) \cos j\omega t
		           +u^o_{ij}(Z) \sin j\omega t] \quad j=1,2,...
\label{seri}
\end{equation} 
Using a perturbation method\cite{zv3}, and taking into account the 
smallness of $h_{ac}$ and $h_{dc}$ 
by neglecting cubic and higher terms in these parameters,
we obtain the leading contribution to $u_i(Z,t)$
(determined by $u_{i0}$, $u^e_{i1}$ and $u^e_{i2}$).
The quantities of interest e.g., the magnetic moment or magnetic 
torque, can be written down\cite{zv3} in terms of the 
surface values of the fields. By  substituting Eq. 
(\ref{seri}) into (\ref{ueq}) and (\ref{bcds}) we get 
\begin{eqnarray}
u_i(Z_s,t)&=&h(t)[\cos \psi \delta_{iX}+\sin \psi \delta_{iY}] \nonumber 
\\
	  &+&\frac{h(t)\left|h(t)\right|}{3 \mu^2}
	  [e_{1i}\cos^2\psi+e_{2i}\cos \psi \sin \psi+e_{3i} \sin^2\psi] 
\qquad i=X,Y,
\label{utot}
\end{eqnarray}
where $\delta_{ij}$ is the Kronecker symbol. The 
angular dependence of the second term, nonlinear and 
nonanalytic in the field,  is 
identical to that found in the case of constant applied field. 
Thus it remains to 
investigate the temporal dependence of various quantities which are
also nonanalytic.
The nonlinear transverse magnetic moment for gaps with nodes
can be  written (Appendix \ref{ap1}) as
\begin{eqnarray}
\label{msep}
m_\bot(\psi,t)&=&\frac{{\cal S} \lambda_{ab}}{6 \mu \pi}  
\frac{H_a(t)\left|H_a(t)\right|}{H_0}
[e_{3X} \sin^3 \psi-e_{1Y}\cos^3 \psi \nonumber \\
&+&\cos \psi \sin\psi ((e_{1X}- e_{2Y}) \cos \psi+
(e_{2X}-e_{3Y}) \sin \psi)]\\
&\equiv& N_m\frac{h(t)\left|h(t)\right|}{h^2} f(\psi) 
\equiv N_m {\cal M}_\bot (\psi,t)  \nonumber
\end{eqnarray}
where ${\cal S}$ is the slab area and the factor $\lambda_{ab}$ reflects
that the nonlinear effects arise from currents
present within a length scale on the 
order of penetration depth.
We have introduced the normalization factor 
\begin{equation}
N_m=\frac{{\cal S} 
\lambda_{ab} (H_{dc}+H_{ac})^2}{6 \mu \pi H_0} \frac{1}{3\sqrt{3}}
\label{nm}
\end{equation} 
and the normalized transverse moment ${\cal M}_\bot $.
The angular
dependence of  $m_\bot(\psi,t)$ or of ${\cal M}_\bot(\psi,t)$
is given  by the function
\begin{equation}
f(\psi)=3\sqrt{3}[e_{3X} \sin^3 \psi-e_{1Y}\cos^3 \psi 
+\cos \psi \sin\psi ((e_{1X}- e_{2Y}+(e_{2X}-e_{3Y})) \sin \psi].
\label{f}
\end{equation} 
This result is independent of time and thus it is the
same as in the static case.
For a pure $d$-wave $f(\psi)$ has a maximum value of  unity,
and so does the normalized moment as a function of angle.

Experimentally, one can best detect the
nonlinear effects by examining the dominant harmonics (time
Fourier components) of $m_\bot(\psi,t)$, which are at 
$2\omega$ and $3\omega$. Analysis of 
the field and angular dependence of these harmonics
makes it possible to accurately determine the position of  the nodes in 
the energy gap. One could obtain the same information
from $m_\|(\psi,t)$,
discussed in Appendix \ref{ap2}, but with greater
difficulty because of the extremely large linear signal.
The angular dependence of  $m_\|(\psi,t)$ is different from that of 
$m_\bot(\psi,t)$ but they both have identical temporal dependence. 

Since the measurements are performed in the Meissner state,
one must have
$H_a(t) \leq H_{f1}$, where $H_{f1}$ is the field of first flux
penetration, somewhat larger than $H_{c1}$. Therefore, one
wishes to optimize the experimental
signal (i.e. the size of the $2\omega$ or $3\omega$ harmonics), 
by an appropriate choice of the 
ac and dc components of the applied field at constant
total maximum field. We
can determine the optimal field mixture i.e., the
ratio $R\equiv H_{dc}/H_{ac}= h_{dc}/h_{ac}$ 
(at {\it fixed} total field $h=h_{dc}+h_{ac} $) which would 
produce the maximum signal for a normalized harmonic, $M_j(\psi)$
\begin{eqnarray}
M_j(\psi)&=&\frac{2}{\pi(1+\delta_{j0})} \int_0^\pi\frac{m_\bot(\psi,t)}{N_m}  
\cos j \omega t d( \omega t)\nonumber \\ 
&=&\frac{2}{\pi (1+\delta_{j0})} \int_0^\pi
\frac{h(t) \left|h(t) \right|}{h^2}  \cos j \omega t d( \omega t) f(\psi)
\equiv \overline{M}_j f(\psi), \qquad j=0,1,2,...
\label{mj}
\end{eqnarray}
which factorizes into field and angular dependent parts,
$\overline{M}_j$ and $f(\psi)$.
This factorization 
is also valid for an anisotropic FS. The field and angular
dependence of $m_\bot$
for an anisotropic FS with nodal lines, as obtained
in Ref. \onlinecite{zv3}, remains separable in a
time varying applied field. As in the static case, FS
anisotropy modifies the specific form of $f(\psi)$ but 
not the field dependence of the coefficient 
$\overline{M}_j$, provided only that the
characteristic field $H_0$ is
properly redefined. Therefore, the determination
of the optimal $R$ discussed below would
remain the same.

Since the angular dependence
$f(\psi)$ is separable, determining the optimal field ratio $R$
reduces to evaluating the coefficient
$\overline{M}_j$, the harmonic of the normalized angular 
amplitude of the nonlinear transverse magnetic moment. The normalizations 
chosen imply that in the static case, $\omega \rightarrow 0$, 
$\overline{M}_0$ is unity.
For $j=2,3$ we obtain from Eq. (\ref{mj})
\begin{mathletters}
\label{m23}
\begin{equation}
\overline{M}_2=\frac{1}{\pi h^2}[h^2_{ac} p+4h_{dc} h_{ac} (\cos p-\frac{\cos 
3p}{3}) 
-(2 h^2 _{dc}+h^2 _{ac})\sin 2p+\frac{h^2_{ac}}{4}\sin 4p]
\end{equation}
\begin{equation}
\overline{M}_3=\frac{1}{\pi h^2}[h^2_{ac}\cos p
-(\frac{4}{3} h^2_{dc}+\frac{2}{3} h^2_{ac}) \cos 3p  
+\frac{h^2_{ac}}{5}\cos 5p-h_{dc} h_{ac}(2 \sin 2p-\sin 4p)]
\end{equation}
\end{mathletters}
where $p\equiv\sin^{-1}(h_{dc}/h_{ac})$.
Results for $\overline{M}_2$ and $\overline{M}_3$ at fixed maximum field 
as a function of the ratio $R$
are given in Fig. \ref{mom}. 
We show results for $H_{dc} < H_{ac}$ ($R<1$) only,
since for larger values of $R$ these harmonics
are considerably smaller. 
The results
indicate that the optimal applied field should 
be a pure ac field if the harmonic at  $3\omega$
is measured, or $H_{ac}\approx H_{dc}$ for the
$2\omega$ harmonic. If one measures these 
harmonics for several values of $R$ close to optimal, their 
dependence on $R$, as we shall see in Section III, can serve to 
experimentally distinguish nodes from small minima in the energy gap.
The reduction of 
the amplitude for $\overline{M}_j$, $j=2,3$ compared to the static case 
(a factor of $\approx 6$) is more than compensated by the advantages of ac 
techniques.\cite{sens} 

We can apply  a similar analysis to the magnetic torque,
${\bf \tau}={\bf m} \times {\bf H}$. Torque
magnetometry is an extremely sensitive experimental
technique, which has already been used to investigate
anisotropic magnetic properties of superconductors at 
higher fields.\cite{kogan,ross,maki} Here we are
interested in  $\tau_z$, which has an angular dependence 
identical to that of $m_\bot(\psi)$. 
The main harmonics arising from the nonlinear
response are at $3\omega$
and $4\omega$.  Using  a convention 
similar to that in Eq. (\ref{mj}) we define
\begin{equation}
T_j(\psi)=\frac{2}{\pi} \int_0^\pi
\frac{\left|h(t)\right|^3}{h^3} \cos j \omega t d(\omega t)  f(\psi)
\equiv \overline{T}_j f(\psi), 
\qquad j=3,4
\label{tj}
\end{equation}
where we make use of the separability of the angular dependence. 
To determine the optimal $R$ we need only consider $\overline{T}_j$, the 
harmonic of normalized angular amplitude of the nonlinear torque.   
Elementary  integration in Eq. (\ref{tj})
yields
\begin{mathletters}
\label{t34}
\begin{eqnarray}
\overline{T}_3&=& \frac{1}{\pi h^3}[\frac{h^3_{ac}}{2}p+3h_{dc} h^2_{ac}\cos p
-(\frac{4}{3}h^3_{dc}+2h_{dc}h^2_{ac})\cos 3p  \\  \nonumber
&+&\frac{3}{5}h_{dc}h^2_{ac}\cos 5p
-(3 h^2_{dc} h_{ac}+\frac{3}{4}h^3_{ac})(\sin 2 p-\frac{\sin 4p}{2})
-\frac{h^3_{ac}}{12}\sin 6p] 
\end{eqnarray}
\begin{eqnarray}
\overline{T}_4&=&\frac{1}{\pi h^3}[\frac{h^3_{ac}\cos p}{2}
-(2h^2_{dc} h_{ac}+\frac{h^3_{ac}}{2})\cos 3p
+(\frac{6}{5} h^2_{dc}h_{ac}+\frac{3}{10}h^3_{ac})\cos 5p  \\  \nonumber
&-&\frac{h^3_{ac}}{14} \cos 7p-\frac{3}{2} h_{dc} h^2_{ac}\sin 2p 
+(h^3_{dc}+\frac{3}{2}h_{dc} h^2_{ac}) \sin 4p
-\frac{h_{dc}h^2_{ac}}{2}\sin 6p]  
\end{eqnarray}
\end{mathletters}
The results for $\overline{T}_j$ as a function of R at 
fixed maximum field,
seen in Fig. 
\ref{torq}, show that for gaps with nodes it is optimal to 
measure the $4 \omega$  time FC at pure ac applied field,
and the $3\omega$ component at $R\approx0.35$.
As in our discussion of $\overline{M}_j$, one can see that
FS anisotropy does not alter $\overline{T}_j$
or the optimal value of $R$.

\subsection{Microstrip Resonator}
The nonlinear aspects we discuss here affect many
other frequency dependent phenomena. 
We refer here briefly as an example to the
nonlinear magnetic response
of a microstrip resonator at low temperature. The 
particular geometry is that from Ref. \onlinecite{scal} where
the nonlinear dynamics was discussed for higher temperatures,
when the relation between ${\bf j}$  and 
${\bf v}$ is analytic. At low temperature the effect is quite
different, as 
the nonlinearities are nonanalytic.
For a cylindrical, isotropic FS, Eq. 
(\ref{nonlinjv}) can be expressed as:
\begin{equation}
{\bf j}=-e \rho {\bf v}+{\bf j}_{qp}=-e \tilde{n}_s {\bf v},
\label{j}
\end{equation} 
where we use a notation distinguishing the electron density $\rho$ from 
the superfluid density tensor $\tilde{n}_s$, which also includes 
anisotropic, nonlinear effects due to quasiparticle excitations. The 
specific form of $\tilde{n}_s$ depends on the angular form of the 
energy gap and can be simply obtained from $j_{qpi}$ in Eq. (A1).
The effects that we investigate can be viewed as arising either 
 from a time dependent applied magnetic field or a driven time varying 
current. Following Ref. \onlinecite{scal}, we consider a pure $d$-wave gap 
with an applied field (or driven current) along nodal and antinodal 
directions. In both cases ${\bf 
j}$ $\|$ ${\bf v}$. We have
\begin{equation}
\frac{\lambda^2(T=0,{\bf j}=0)}{\lambda^2(0,{\bf j})} = 
\frac{n_s(0,{\bf j})}{n_s(0,0)}=1-b_\psi \frac{\left| {\bf j}\right|}
{\left|j_c\right|}
\label{pen}
\end{equation} 
where $\psi=0$ corresponds (Fig. \ref{axes}) to ${\bf j}$ along the 
nodal and $\psi=\pi/4$ to ${\bf j}$ along the antinodal direction.  
 From Appendix A we obtain $b_{\psi=0}=1/2$ 
and $b_{\psi=\pi/4}=1/2\sqrt{2}$.
The critical current is $j_c=n_s(0,0)e \Delta_d/v_f$ and 
in the absence of quasiparticle excitations at $T=0$, $j=0$, the 
superfluid density $n_s(0,0)$ equals
$\rho$.

We next examine the inductance per unit length, $L$, and show how
the low temperature form of ${\bf j(v)}$ affects its nonlinear 
corrections. These 
considerations are then readily applied to other quantities of interest 
such as the resistance per unit length or the quality factor $Q$.
We use the expression for the
 inductance\cite{scal}
\begin{equation}
L=\frac{\int_S (H^2+\lambda^2(0,j)(4\pi j/c)^2)dS}{4\pi(\int_S jdS)^2}
\label{l}
\end{equation}
with the integration being over a cross section of the microstrip having 
width $w$ and thickness $d$. This expression can be rewritten as
\begin{equation}
L= L_0+\Delta L\frac{\left|I\right|}{\left|I_c\right|}
\label{l2}
\end{equation}
where $L_0$ is the inductance in the linear response, $I=\int_S jdS$ and 
$I_c\int_S j_c dS= w d j_c$.
We make the approximation\cite{scal} that the total correction
$\Delta L$ due to 
the nonlinear response can be replaced by the kinetic part only, 
$\Delta L_{KI}$. Then, by using Eqs. (\ref{pen}), (\ref{l}), and (\ref{l2}) we 
obtain $L \approx L_0+\Delta L_{KI} \left|I\right|/\left|I_c\right|$
with 
\begin{equation}
 \Delta L_{KI}=\frac{4 \pi b_\psi \lambda^2(0,0) w d \int_S \left| j^3 \right| 
dS} {c^2 \left|\int_S jdS\right|^3}.
\label{l3}
\end{equation}
One can consider also  the
frequency dependent $Q$ or the resistance, 
which can be decomposed into linear and nonlinear parts as in Eq. 
(\ref{l2}). The power law behavior of the nonlinear corrections  in these 
quantities i.e, linear in $I/I_c$, (instead of  $(I/I_c)^2$ at higher 
temperatures) will affect generation of higher harmonics. 
By using the transmission line equations\cite{scal} 
for a microstrip, one easily sees that
these two types of nonlinearity give
rise to second harmonic generation at low $T$ eventually
crossing over to third harmonic at 
higher temperatures. 
\section{Detection of Nodeless Gap Functions}
In this section we investigate the quasi-static nonlinear magnetic response of 
anisotropic  energy gaps with ``quasinodes''  rather than with nodes. The
nonlinear electromagnetism samples the pairing state in the
bulk. If the nodes 
are absent  only in a layer of 
thickness much smaller than $\lambda$
near the surface\cite{bound,green}, our results from Section 
\ref{harm}
apply unmodified.
We consider here a typical candidate, $d+is$,\cite{sid,kot,joy} 
energy gap with the $d$ component being
dominant. The results presented are quite similar to
those that can be obtained for  other nodeless 
gaps such as $d_{x^2-y^2}+id_{xy}$\cite{lau,bal,movs} or an anisotropic 
$s$-wave. Generally, in the absence of nodes,
the analysis of the low frequency nonlinear magnetic 
response is more complicated.
There are threshold effects that decisively affect the field
dependence.

We can approximate a $d+is$\cite{kot,joy} gap near its minima by
\begin{equation}
\left| \Delta(\phi_n) \right|
\approx (\mu^2 \Delta^2_d \phi^2_n+\Delta^2_s)^{1/2}, \qquad n=1,2,.. 
\label{dis}
\end{equation}
with $\Delta_s \ll \Delta_d$ and $\mu=2$ for the usual form of a 
$d$-wave OP. The small minima $(\Delta_s)$ are located at the positions 
of the nodal points in a $d$-wave gap. 
The current due to the quasiparticle excitations is obtained from Eq. 
(\ref{jq}) 
\begin{equation}
{\bf j}_{qp}{\bf (v)}
\approx-\frac{e}{2} \sum_n N_f v_f \hat{x}_n 
\frac{v_f^2({\bf v} \cdot \hat{x}_n)^2-\Delta^2_s}{\mu^2 \Delta_d^2} 
\Theta(v_f^2({\bf v} \cdot \hat{x}_n)^2-\Delta^2_s) 
\qquad n=1,2,...
\label{jqi}
\end{equation}
where the step function arises from  the phase space 
for allowed quasiparticle excitations  
($\left| \Delta(\phi_n) \right |+{\bf v}_f \cdot {\bf v} < 0$ ),  given by
\begin{equation}
\phi^2_n \leq                          
\frac{v_f^2({\bf v} \cdot \hat{x}_n)^2-\Delta^2_s}{\mu^2 \Delta_d^2},
\label{phi}
\end{equation}
which is reduced compared to the  $d$-wave gap. The step functions result
in a threshold effect: it now takes a minimum field to create 
quasiparticles.  However, as we 
shall later see, a small admixture of $s$-wave in a $d+is$ gap can enhance 
{\it the maximum value} of the
nonlinear effects at fields above threshold. From Eq. (\ref{phi}) it follows 
that the nonlinear 
effects are absent whenever $H_a(t)$ is small enough so that
$\left| v \right| < v_T \equiv \Delta_s/v_f$.
In dimensionless form the threshold field $h_T$ and the threshold 
velocity $u_T$ are
\begin{equation}
h_T\equiv \delta, \qquad u_T\equiv \frac{\delta}{\mu}, \qquad
\delta\equiv\frac{\Delta_s}{\Delta_d}.
\label{hut}
\end{equation} 
Substitution of Eq. (\ref{jqi}) into (\ref{lap}) yields
\begin{equation}
\partial_{ZZ}u_i-u_i
+{\rm sgn}(h(t))(u_i^2-u^2_T)\Theta(\left|u_i\right|-u_T)=0, \qquad i=X,Y,
\label{ueqi}
\end{equation}
where $u_i$ is given by Eq. (\ref{units}) with $\Delta_{eff}=\Delta_d$.
By modifying the methods used  in this case for a static 
field\cite{zv3} to include the time dependence
we obtain
\begin{eqnarray}
u_i(Z_s,t)&=&h(t)[\cos \psi \delta_{iX}+\sin \psi \delta_{iY}]
	  +[\frac{1}{3\mu^2}h(t)\left|h(t)\right|
(\cos^2 \psi \delta_{iX}+\sin^2 \psi \delta_{iY})-u^2_T {\rm sgn}(h(t))
 \nonumber \\ &+&\frac{4 u_T^3}
{3 h(t)(\cos \psi \delta_{iX}+\sin \psi \delta_{iY})}]
 \Theta(\left|h(t)\right|-
\frac{h_T}{(\cos \psi \delta_{iX}+\sin \psi 
\delta_{iY})}), \quad i=X,Y
\label{utoti}
\end{eqnarray} 
for the solution at the surface of a slab, $Z=Z_s$ $(Z_s \gg 1)$, which 
suffices to express the quantities $m_\bot$ and  $\tau_z$.  
Nonlinear effects are present only for 
$h(t) \geq h_{Ti}(\psi,\delta) \geq h_T$, where 
$h_{TX}\equiv\delta/ \cos \psi$ and  
$h_{TY}\equiv\delta / \sin \psi$ are the angle dependent threshold 
fields required to excite jets of quasiparticles centered along the $X$ and 
$Y$ axes respectively.
The transverse magnetic moment is
\begin{eqnarray}
\label{mdis}
m_\bot(\psi,t)&=&\frac{{\cal S} \lambda_{ab}}{6 \mu \pi}
\:[\frac{H_a(t)\left| H_a(t) \right|}{H_0} 
\cos\psi \sin \psi[\cos \psi \Theta(\left| h(t) \right|-h_{TX})
-\sin \psi \Theta(\left| h(t) \right|-h_{TY})]  \nonumber \\
&+&3\delta^2 H_0 \: {\rm sgn}(H_a(t))
[\cos \psi \Theta(\left| h(t) \right|-h_{TY})
-\sin \psi \Theta(\left| h(t) \right| -h_{TX})] \\
&-&2\delta^3  \frac{H_0^2}{H_a(t)}
[\cot \psi \Theta(\left| h(t) \right| -h_{TY}) 
-\tan \psi \Theta(\left| h(t) \right|-h_{TX})]], \nonumber
\end{eqnarray}
where we see that, as stated above,
the temporal and angular dependence of $m_\bot(t,\psi)$ are no 
longer separable. To investigate the behavior of this expression at 
different fields and admixtures of $s$-wave OP it is useful to 
employ the scaling relation:
\begin{equation}
m_\bot(\kappa h(t), \kappa \delta)=\kappa^2 
m_\bot(h(t),\delta)
\label{scal}
\end{equation}
valid for any angle $\psi$, at fixed field ratio, $R$. 
The quantity $\kappa$ is an arbitrary scaling factor.

Again, we investigate the dominant harmonics
of $m_\bot(\psi,t)$. Because of the
non-separability of temporal and angular dependences,
one cannot introduce $\overline{M}_j$
as in Eq. (\ref{mj}). One
must now carefully consider both the 
harmonics
\begin{equation}
M_j(\psi)= \frac{2}{\pi}\int^\pi_0 
\frac{m_\bot(\psi,t)}{N_m} \cos j \omega t 
d(\omega t), \qquad j=2,3
\label{m3i}
\end{equation}
and the relevant angular Fourier components (FC's)
in terms of which the angular dependence can be
analyzed. The most important angular FC 
is that which reflects the main angular periodicity, $\pi/2$ of the 
energy gap
\begin{equation}
M_{j}^4= \frac{8}{\pi}\int^{\pi/4}_0 M_j(\psi) \sin 4\psi d\psi, \qquad j=2,3
\label{m3ij}
\end{equation}
We will focus in  the rest of this section on the dominant
response of the nonlinear magnetic moment
at $2 \omega$ and $3 \omega$. 
 
In Fig. \ref{mom3} we show our results for the angular 
dependence of the $3 \omega$ harmonic, 
$M_3(\psi)$. The details of its calculation are given in
Appendix \ref{ap3}. We display the range $\psi \in [0,\pi/4]$.
We have taken $R=0$ (guided by  Fig. \ref{mom}) and a
dimensionless field amplitude
$h=h_{ac}=0.05$. 
This value, using typical material
parameters for YBCO, corresponds to $H_{ac} \approx H_{f1}$.  
The solid line represents the $d$-wave result, normalized such 
that the maximum of ${\cal M}_\bot(\psi)$ for the static case
is unity. 
This normalization is employed throughout this section. 
The broken lines are labeled by the corresponding ratios
$\delta$.

 An obvious feature seen in Fig. \ref{mom3}, is the  
enhancement of the maximum value of the
signal that may arise from small admixtures of an $s$-wave OP.
At fixed field, the maximum enhancement 
of $M_3(\psi)$ is much more pronounced than the moderate one for 
$m_\bot(\psi)$ found in the static case (Fig. 7 in Ref. 
\onlinecite{zv3}). A similar enhancement 
is also present in $M_2(\psi)$. This result is rather 
unexpected since such an admixture would reduce the available 
phase space (recall Eq. (\ref{phi})) for quasiparticle excitations. 
This enhanced signal is
of non-negligible significance in planning experiments. We therefore
pause here to explain its physical origin.

The nonlinear transverse magnetic moment arises  from 
contributions due to components of the quasiparticle 
current ${\bf j}_{qp,n}{\bf (v)}$ at different nodes (quasinodes).
These contributions partly cancel each other as each node,
in effect, tries to twist the magnetic moment away from itself.
Thus, if a magnetic field is 
applied along an antinode, the cancelation of such components 
is complete and 
$m_\bot=0$. To explain the peculiar enhancement in $M_3(\psi)$, 
we need to investigate how the introduction of $\Delta_s$ increases 
the {\it asymmetry} of such contributions from  different jets. The
asymmetry, which is reflected in the angular asymmetry of
the curves in Fig. \ref{mom3} for $\delta>0$, is brought about
by one of the
nodes being above its angular dependent
threshold field (see below Eq. (\ref{utoti})) while the other is still below. It
was seen in Ref. \onlinecite{zv3} that this was
responsible for the small enhancement seen in the dc case. Now we
will see quantitatively how the larger enhancement in the ac case arises
  from the same source.

We denote the contributions to  ${\cal M}_\bot \equiv m_\bot/N_m$
of the individual jets along the $X$ and $Y$ 
axes at a fixed angle $\psi_0$ by 
${\cal M}_\bot(\psi_0,t)_X$, ${\cal M}_\bot(\psi_0,t)_Y$:
\begin{equation}
{\cal M}_\bot(\psi_0,t)={\cal M}_\bot(\psi_0,t)_X+{\cal M}_\bot(\psi_0,t)_Y.
\label{xy}
\end{equation} 
In Fig. \ref{momx}, computed for the same fields
as in Fig. \ref{mom3}, we show
${\cal M}_\bot(\psi,t)_X$, ${\cal M}_\bot(\psi,t)_Y$ and their
sum, ${\cal M}_\bot(\psi,t)$,  as a function of time $\omega t$
at a fixed angle $\psi\equiv \psi_0\equiv 25^0$ in the region where the
maximum of $M_3(\psi; \delta=0.02)$ occurs. 
We see in Fig. \ref{mom3} that 
the enhancement in this case is large, a factor of about two 
over the $d$-wave signal.
The quantities in the figure are shown
 for both $\delta=0$ and $0.02$. The curves are
labeled by two indices, the first a letter denoting
the jet, or sum of jets, plotted, and the second being the
value of $\delta$.
The range $\omega t \in [0,90^0]$
shown in the figure could 
be extended over the whole period by symmetry.
For the chosen value of $\psi_0$, the threshold fields satisfy  
$h_{TX}(\psi_0) < h_{TY}(\psi_0)$ and consequently 
the magnitude of the jet along the $X$ axis is 
greater than that  of the one along the $Y$ axis. 
With the admixture of an $s$-wave OP, there is a region, $\omega t 
\lesssim 
20^0$, where the magnitude of ${\cal M}_\bot(\psi_0,t)_X$ is reduced 
{\it less}
than that of ${\cal M}_\bot(\psi_0,t)_Y$. Therefore the cancelation of the 
contributions from the two different jets 
takes place to a 
lesser
extent, resulting in an overall increase of ${\cal M}_\bot(\psi_0,t)$. 
However this  accounts for only part of the enhancement 
 seen in Fig. \ref{mom3}, basically the part found in the dc case,
which corresponds to that at $\omega 
t=0$  in the Figure. 
The remaining part of the increase in $M_3(\psi)$ is explained by the 
convolution of the time dependence of ${\cal M}_\bot(\psi,t)$ and $\cos 3 
\omega t$ in the definition of $M_j(\psi)$,
Eq. (\ref{m3i}). 
When the curves representing  $m_\bot(\psi,t)$ for $\delta=0, 0.02$ 
and $\cos 3 \omega t$, (which is also shown in Fig. \ref{momx})
are multiplied 
the results obtained, which are the integrands 
 in Eq. (\ref{m3i}), are as shown
in Fig. \ref{momt}. The solid curve represents the
pure $d$-wave result and the broken curve that for $\delta=0.02$. 
The ratio of the two net areas between the curves shown and
the horizontal axis is about a factor of two, i.e. the factor 
between $M_3(\psi_0)$ for $\delta=0$ and for $\delta=0.02$ 
in Fig. \ref{mom3}. Thus, the enhancement is
explained.

It is straightforward to perform a very
similar analysis for $M_2(\psi)$ using its 
explicit expression from Appendix \ref{ap3}. From
parity considerations, similar 
to  those made for gaps with nodes, $M_2(\psi)$
vanishes  identically as a function of  $\delta$ 
and $\psi$ for $h_{dc}=0$. In this case, one
chooses $R\approx 0.5$ (see Fig. \ref{mom}) and again finds
an enhanced signal, as in Fig. \ref{mom3}. The size of the
enhancement is largely due to the convolution with
$\cos 2\omega t$.

The non-separable temporal and angular dependences of $m_\bot(\psi,t)$ 
require additional care in trying to determine  the optimal
mixture of dc and ac fields in $H_a(t)$ which maximizes
the harmonics and angle FC's of the signal. From 
the expressions for
$M_{2}(\psi)$ in Eq. (\ref{m23i}) we have obtained
results for its $\pi/2$ angular FC.
We denote this quantity by $M_{2}^4$, as defined
in Eq. (\ref{m3ij}). We show typical
results in Fig. \ref{opt2}. There we plot 
$M_{2}^4$ as a function of $R$ at $h=0.05$, for several values of 
$\delta$. The solid line represents the pure $d$-wave and the broken 
lines are labeled by their value of $\delta$. We see that 
the optimal field ratio depends on the particular $s$-wave admixture. 
Through the scaling relation in Eq. (\ref{scal}) these results can be 
simply extended to other values of $h$ and $\delta$.    
In Fig. \ref{opt3}, we plot $M_{3}^4$ (i.e. the $\pi/2$ angular
FC of the signal at $3\omega$) as a 
function of $R$ at maximum dimensionless field $h=0.05$ for the same values 
of $\delta$. In this case $R=0$ produces the maximal signal, for all  $\delta$. 

It is instructive, in order to learn how to distinguish nodes from quasinodes, 
to contrast Fig. \ref{mom}, which displays $\overline{M}_j$, $j=2,3$,
the angular amplitude of the $2\omega$ and $3\omega$ harmonics 
for a gap with nodes, with Figs. \ref{opt2} and \ref{opt3}.
Because of the separable temporal and angular dependence of $m_\bot(\psi,t)$
when nodes are present, 
the solid lines (pure $d$-wave case)
in Fig. \ref{opt2}, \ref{opt3} are simply proportional 
(with a normalization factor\cite{norm} $\approx$ 1) to the curves 
$\overline{M}_2$ or $\overline{M}_3$ respectively.  
Experimentally, a sensitive test would be to measure $M_{2}^4$, 
$M_{3}^4$ for 
several values of $R$, and compare the $R$ dependence with the corresponding 
solid lines in Fig. \ref{opt2}, \ref{opt3}. The deviation from these lines 
would indicate absence of nodes in the gap.
A further way of analyzing such data is suggested
by Figs. \ref{del2} and \ref{del3} where we show 
$M_{2}^4$ and $M_{3}^4$ respectively as
functions of $\delta$, for several values of $R$ (additional 
results are easily obtained using Eq. (\ref{m3ij}) and the expressions 
in Appendix C). The curves plotted are labeled by their
values of $R$ and are given at a fixed maximum field
$h=0.05$. To compare experimental data with an assumed $d+is$ 
gap, and to determine the $s$-wave admixture, one would
look for a value of $\delta$ which agrees with  a  particular dependence 
of the  measured $M_{2}^4$ or  $M_{3}^4$ at different $R$. 
In these figures we  notice again the implications of the non-monotonic 
behavior of $m_\bot$ (or ${\cal M}_\bot$) with $\delta$. Just as we
found above for the harmonics, there is also a 
substantial enhancement for the angular FC at given $R$ as compared to the 
$\delta=0$ value. It has the same origin, as shown in Fig. 
\ref{momx}, \ref{momt}, predominantly due to the 
convolution of   $M_{j}(\psi, t)$ and $\cos(j \omega t)$, 
$j=2,3$. In Figs. \ref{del2} and \ref{del3} we see that maximum 
such enhancement (at different values of $R$) is located typically in the range 
$0.015 \lesssim  \delta \lesssim 0.025$ at $h=0.05$. More generally, using the 
scaling relation Eq. (\ref{scal}), for an arbitrary maximum field 
$h$ (within the Meissner regime) the maximum signal for $M_{j}^4$, 
$j=2,3$ is 
approximately in the  range $h/3 \lesssim \delta \lesssim h/2$.

Thus, there are several distinct features for which the results
for a nodeless gap with small minima
differ considerably from those
for gaps with true nodes.
These results suggest how to perform  measurements and analyze the 
nonlinear quantities in the low frequency, low temperature regime. 
They can be supplemented with the results from the static case. 
It would also be useful to study the field dependence of $m_\bot$ 
and its Fourier components at fixed $R$ while changing the maximum field 
($H_{dc}+H_{ac})$.

Using the methods that we have presented, similar results for
the behavior of the nonlinear magnetic torque can be obtained.
Analogously, results for other gapped states such as $d_{x^2-y^2}
+id_{xy}$ can be derived without difficulty.

\section{conclusions}
We have examined properties of the nonlinear magnetic response of an 
unconventional superconductor at low frequencies. There are two
key points to our
conclusions. First, these frequency dependent nonlinear phenomena
provide a particularly powerful tool to distinguish nodes from
quasinodes. Second, the experimental sensitivity  that allows one to
perform ``node spectroscopy'' can be considerably enhanced over the
corresponding dc result.

For gap functions 
with nodal lines the nonlinear moment ($m_\bot$, $m_\|$) and torque 
($\tau_z$) have separable field (and hence, time) and angular dependences. 
These quantities are respectively quadratic and cubic (up to
a sign determined by parity) in the instantaneous 
applied field. The nonlinear magnetic response for nodeless gaps is 
more complicated: the time (or field) and angular dependences
are not separable and the 
field dependence is not a simple power law. We have discussed how these 
features can be studied by measuring higher harmonics (referred
to the frequency of the applied ac field) of various physical 
quantities. 
For an applied field $H_a(t)=H_{dc}+H_{ac} \cos \omega t$, at a fixed 
maximum $H_{dc}+H_{ac}$  we have computed the optimal 
ratio $R\equiv H_{dc}/H_{ac}$ at which the signal (for a certain higher 
harmonic) is largest. For gaps with lines of nodes it 
is most favorable to measure the $2 \omega$ harmonic of $m_\bot$ at 
$R \approx 0.5$ 
or the $3 \omega$ harmonic at $R=0$, while for $\tau_z$  the 
best alternatives are the $3 \omega$ 
harmonic 
at $R\approx 0.35$ and the $4 \omega$ harmonic
component at $R=0$. These 
results are independent of the specific angular dependence of the 
gaps with nodal lines and are also valid for an anisotropic FS. In the absence 
of nodes, as we illustrate through the 
example of a $d+is$ gap, selecting the optimal value of $R$ to maximize 
higher harmonics is less simple but more 
rewarding. The result depends on the particular 
admixture of $s$-wave component to the OP and hence provides
for the quantitative determination of this admixture.
The results we shown are sensitive to the angular position of nodal 
(quasinodal) lines, allowing for detailed characterization of the 
minima in the energy gap.

The analysis presented here can be generalized to 
include point nodes, and to triplet pairing. The field 
dependence for gaps with point nodes is different 
than that found for gaps with lines of nodes. For example, $m_\bot$ is 
cubic rather than quadratic in the instantaneous value of the applied  field. 
Using these considerations  and the temperature range attainable by a dilution 
refrigerator it should be feasible to experimentally study other candidates for 
unconventional  superconductivity, such as Heavy Fermions, Organic 
Superconductors and ${\rm SrRu_2O_4}$. Analysis of the material parameters 
(to estimate $H_0$ and $H_{f1}$) 
shows that in many cases
the expected signal should be at least as large as
for HTSC's. 
Experiments 
investigating magnetic response in these materials, at sub-Kelvin 
temperatures, are currently under considerations.\cite{gold} 

\acknowledgments
We are grateful to  A. Bhattacharya and A. M. Goldman for many useful 
conversations related to the experimental work and techniques to 
measure  the nonlinear magnetic moment. We also thank J. A. 
Sauls for discussions. I. \v{Z}. acknowledges 
support from the University of Minnesota, Graduate School Doctoral 
Dissertation Fellowship.

\appendix
\section{Currents}
\label{ap1}
Integration of Eq. (\ref{jq}) without $a-b$ plane FS anisotropy 
yields\cite{zv3}
\begin{equation}
j_{qpi}=\frac{1}{2}e N_f \frac{v_f^3}{\mu \Delta_{eff}}[e_{1i}(\psi)
v_{X}^2+e_{2i}(\psi) v_{X} v_{Y}+e_{3i}(\psi) v_{Y}^2], \qquad
i=X, Y, \label{jqds}
\end{equation}
where the coefficients $e_{1,2,3\:i}(\psi)$ are defined for
various ranges of $\psi$. Depending on the particular orientation of 
${\bf H}_a$, as given by the angle $\psi$, these coefficients describe 
contributions from different jets. The results are quadratic in the superfluid 
velocity. For a $d\pm s$ OP we obtain:  
\begin{mathletters}
\label{es}
\begin{equation}
e_{1X}=e_{3Y}=c^3-s^3, \:
e_{2X}=2 e_{3X}=2 e_{1Y}=e_{2Y}=-2 cs(c-s),  \: \: \: \psi \in [\nu,
\frac{\pi}{2}-\nu]
\end{equation}
\begin{equation}
e_{1X}=-e_{3Y}=-(c^3+s^3), \:
e_{2X}=-2e_{3X}=2e_{1Y}=-e_{2Y}=2 cs(c+s), \: \: \: \psi \in
[\frac{\pi}{2}-\nu, \pi+\nu].
\end{equation}
\end{mathletters}
where $c \equiv \cos \nu$, and $s \equiv \sin \nu$. The pure $d$-wave 
limit corresponds to $c=1$ and $s=0$.
\section{Magnetic Moment for a gap with lines of nodes}
\label{ap2}
The magnetic moment can be calculated using 
only the surface values of the fields.\cite{jcp,zv3,jap}  For the geometry
considered here, the  magnetic moment, 
\begin{equation}
{\bf m}= \frac{1}{2 c} \int d {\bf r} {\bf r} \times {\bf j(v)},
\label{moment}
\end{equation}
can be expressed as:
\begin{equation}
m_{x,y}=-\frac{{\cal S} d \:H_{a\:x,y}} {4 \pi} \mp
\frac{{\cal S} c}{2 \pi e} v_{y,x}(d/2), \label{msur}
\end{equation}
where $x$ and $y$ are orthogonal axes fixed in space, ${\cal S}$ is the 
slab area, and we have used that ${\bf v}$ is odd in $z$.
For a slab, $(d\gg \lambda_{ab})$, which is the case of experimental
interest,\cite{sv} the nonlinear magnetic moment can be obtained from 
$u_i(Z,t)$, given in Eq. (\ref{utot}).  For a gap with
node lines it is $\propto$ $h(t)\left|h(t)\right|$.
The transverse component, $m_\bot=m_{X} \cos \psi +m_{Y} \sin \psi$, 
perpendicular to the applied field, is obtained from Eq. (\ref{msep})
and the longitudinal component of the nonlinear magnetic moment,  
$m_\|= -m_X \sin \psi + m_Y\cos \psi$, is given by
\begin{eqnarray}
\label{mlo}
m_\|(\psi)&=&\frac{{\cal S} \lambda_{ab}}{6 \mu \pi}  
\frac{H_a(t)\left|H_a(t)\right|}{H_0}
[ e_{1X} \sin^3 \psi+e_{3Y}\cos^3 \psi \\
&+&\cos \psi \sin\psi ((e_{2X}+ e_{1Y}) \cos \psi+
(e_{3X}+e_{2Y}) \sin \psi)]. \nonumber
\end{eqnarray}
For a pure $d$-wave OP and $\psi \in [0,\pi/2]$, the angular 
dependence of $m_\bot$ 
(its normalization as given in  Eq. (\ref{f})) is
$f(\psi)=3 \sqrt{3}\cos\psi \sin\psi [ \cos\psi-\sin\psi]$
and that for $m_\|$ it is $f_\|(\psi)= 3 \sqrt{3}[\cos^3 \psi+\sin^3 \psi]$. 
While the angular modulation for $m_\|$ is somewhat smaller than that 
of $m_\bot$, the maximum of $m_\|$ is enhanced by a factor of 
$3\sqrt{3}$. This enhancement combined with $ac$ measuring 
techniques\cite{sens} could facilitate examining the higher 
harmonics due to the nonlinear response.
\section{Fourier Components For A Nodeless Gap}
\label{ap3}
We give here the analytic expression for the harmonics 
$M_{j}(\psi)$,
valid  for a $d+is$ gap and $h_{dc}<h_{ac}$. Integration of Eq. 
(\ref{m3ij}) and substitution of $m_\bot$ from Eq. (\ref{mdis}) yields
\begin{eqnarray}
\label{m23i}
M_j(\psi)&=&\frac{2}{\pi}\frac{3 \sqrt{3}}{h^2}[\cos\psi \sin\psi
[\cos \psi (p_{1j}\Theta_1+p_{3j}\Theta_3)
-\sin \psi (p_{2j}\Theta_2+p_{4j}\Theta_4] \nonumber \\
&+&3\delta^2
[-\sin \psi (q_{1j}\Theta_1+q_{3j}\Theta_3)
 +\cos \psi (q_{2j}\Theta_2+q_{4j}\Theta_4] \\
&-&2\delta^3
[-\tan \psi (r_{1j}\Theta_1+r_{3j}\Theta_3)
 +\cot \psi (r_{2j}\Theta_2+r_{4j}\Theta_4]], \qquad j=2,3 \nonumber
\end{eqnarray}
where we have introduced the abbreviations
\label{theta}
\begin{equation}
\Theta_{1,2} \equiv 
\Theta[a+b-h_{TX,Y}(\psi, 
\delta)], \quad
\Theta_{3,4} \equiv 
\Theta[b-a-h_{TX,Y}(\psi, \delta)],
\end{equation}
and $a$, $b$ are replacing $h_{dc}$ and  $h_{ac}$ respectively. Since we work 
at a fixed maximum field $(h_{dc}+h_{ac}=const.)$, 
and  $R=h_{dc}/h_{ac}$, $R\in [0,1)$, 
we set $a=(h_{dc}+h_{ac})R/(1+R)$, $b=(h_{dc}+h_{ac})/(1+R)$.
The coefficients in Eq. (\ref{m23i}) for the $2 \omega$ component are
\begin{mathletters}                              
\begin{equation}
p_{i2}=\frac{b^2}{2} w_i+2ab \sin w_i+(a^2+\frac{b^2}{2})\sin 
2w_i+\frac{2}{3} \sin 3w_i+\frac{b^2}{8} \sin 4w_i, \quad i=1,2
\label{pi2}
\end{equation}
\begin{equation}
p_{k2}=\frac{b^2}{2} (w_k-\pi) +2ab \sin w_k+(a^2+\frac{b^2}{2})\sin 
2w_k+\frac{2}{3} \sin 3w_k+\frac{b^2}{8} \sin 4w_k, \quad k=3,4
\label{pk2}
\end{equation}
\begin{equation}
q_{i2}=\sin 2w_i, \qquad i=1,..,4
\end{equation}
\begin{equation}
r_{i2}=\frac{4}{b}\sin w_i
-\frac{4a}{b^2}w_i+\frac{2(2a^2/b^2-1)}{(b^2-a^2)^{1/2}} 
\ln \left|\frac{(b-a) \tan\frac{w_i}{2}+(b^2-a^2)^{1/2}}
	       {(b-a) \tan\frac{w_i}{2}-(b^2-a^2)^{1/2}} \right|, 
\quad i=1,..,4
\end{equation}
\end{mathletters}                              
Those for the $3 \omega$ component are
\begin{mathletters}                              
\begin{equation}
p_{i3}=\frac{b^2}{2} \sin w_i+ab \sin 2 w_i
+\frac{2a^2+b^2}{3}\sin 3w_i
+\frac{ab}{2} \sin 4w_i+\frac{b^2}{10} \sin 5w_i, \quad i=1,..,4
\end{equation}
\begin{equation}
q_{i3}=\frac{2}{3} \sin 3w_i, \qquad i=1,..,4
\end{equation}
\begin{eqnarray}
r_{i3}&=&\frac{2}{b^3}[(4a^2-b^2)w_i -4ab \sin w_i+b^2 \sin 2w_i] \\
&+&\frac{2a(-4a^2+3b^2)}{b^3(b^2-a^2)^{1/2}} 
\ln \left|\frac{(b-a) \tan\frac{w_i}{2}+(b^2-a^2)^{1/2}}
	       {(b-a) \tan\frac{w_i}{2}-(b^2-a^2)^{1/2}} \right|, 
\quad i=1,2 \nonumber
\end{eqnarray}
\begin{eqnarray}
r_{k3}&=&\frac{2}{b^3}[(4a^2-b^2)(\pi-w_k) +4ab \sin w_k-b^2 \sin 2w_k] \\
&-&\frac{2a(-4a^2+3b^2)}{b^3(b^2-a^2)^{1/2}} 
\ln \left|\frac{(b-a) \tan\frac{w_k}{2}+(b^2-a^2)^{1/2}}
	       {(b-a) \tan\frac{w_k}{2}-(b^2-a^2)^{1/2}} \right|, 
\quad k=3,4 \nonumber
\end{eqnarray}
\end{mathletters}                              
where the $w_{1,2,3,4}$ denote possible values 
of $\omega t$ for which 
$\left| h(t) \right| = h_{TX,Y}$. The integral in Eq. 
(\ref{m3i}) can be replaced by integrals
over intervals $\omega t \in [0, w_i]$, $i=1,2$  and $\omega t \in [w_j, 
\pi]$, $j=3,4$. For example, $w_1$ and $w_3$ are the limits  
for contribution of quasiparticle jets along $X$ axis and  $w_2$ and 
$w_4$ those for jets along the $Y$ axis.
They can be expressed as
\begin{equation}
w_{1,2}=\cos^{-1}[(\frac{h_{TX,Y} -a}{b}-1)\: \Theta_{1,2}+1], 
\quad w_{3,4}=\cos^{-1}[(\frac{-h_{TX,Y} -a}{b}-1) \: \Theta_{3,4}-1],
\label{wi}
\end{equation}
for a pure $d$-wave gap, $w_i=\cos^{-1}(-a/b)$, $i=1,2,3,4$. 
The previously given expressions for $M_{j}(\psi)$, $j=2,3$, 
can be then used 
(see Eq. (\ref{m3ij})) to 
obtain $M_{2}^4$, $M_{3}^4$, the $\pi/2$ 
angular 
Fourier components of the $2\omega$  and $3\omega$
signals of the transverse magnetic moment.

\begin{figure}
\caption{Coordinates and  definitions used in the paper. 
The FS and the  energy gap are shown
schematically. The crystallographic directions $a$ and $b$
are indicated. The  orthogonal $X$ and $Y$ axes 
are  along the nodal directions of the pure 
$d$-wave gap. The $d+s$
nodal directions, labeled 1,2,3,4 are shifted by an angle $\pm \nu$
(see Eq. (\protect\ref{nu})) from their $\Delta_s=0$ values. 
The applied magnetic field, ${\bf H}_a$, forms an angle $\psi$ with the
$+Y$ direction.}
\label{axes}
\end{figure}
\begin{figure}
\caption{ Harmonics, $\overline{M}_j$, $j=2,3$, at 
$2 \omega$, $3\omega$ of the  normalized  
transverse magnetic moment  
angular amplitude (Eq. (\protect\ref{mj})), plotted as a function of 
the field ratio,  
$R\equiv h_{dc}/h_{ac}$.
Normalization is taken so that 
in the static case, $\omega \rightarrow 0$, $\overline{M}_0$ is unity.}
\label{mom}
\end{figure}
\begin{figure}
\caption{Harmonics, $\overline{T}_j$, $j=3,4$, at 
$3 \omega$, $4\omega$ of  the normalized  magnetic torque 
amplitude,
(Eq. (\protect\ref{tj})),   
as a function of
$R\equiv h_{dc}/h_{ac}$. Results are shown for gaps with lines 
of nodes. Normalization is as in Fig. \protect\ref{mom}.}
\label{torq}
\end{figure}
\begin{figure}
\caption{Angular dependence of $M_3(\psi)$, 
(see Eq. \protect{\ref{m3i}}), the $3 \omega$ harmonic
of the normalized transverse magnetic moment, ${\cal M}_\bot$, 
for various admixtures of $s$-wave in an $d+is$ energy gap. Results are 
shown at $h=h_{ac}=0.05$ 
and normalized (as in Fig. \protect\ref{mom}) 
so that the maximum of ${\cal 
M}_\bot$ for a pure $d$-wave gap is unity. Lines are labeled by the
ratio of $s$- and $d$-wave amplitudes, $\delta\equiv 
\Delta_s/\Delta_d$.}
\label{mom3}
\end{figure}
\begin{figure}
\caption{Time dependence in a $d+is$ state of 
${\cal M}_\bot(\psi,t)$ (Eq. (\protect\ref{xy})) 
and its contributions 
 from jets along the $X$ (${\cal M}_\bot(\psi,t)_X$) and $Y$
(${\cal M}_\bot(\psi,t)_Y$) axes. Normalization and field are as in
Fig. \protect\ref{mom3}. Results shown are at
fixed angle $\psi_0=25^0$ for which (Fig. \protect\ref{mom3}) 
 the signal  for $\delta=0.02$ is substantially enhanced
as compared to the pure $d$-wave case. 
Curves are labeled by the axes describing the
contribution from a particular jet, and the value of $\delta$. 
Also shown is $\cos 3 \omega t$, which enters the 
definition of $M_3(\psi)$ in Eq. (3.9) and accounts for part of 
the enhancement.}
\label{momx}
\end{figure}
\begin{figure}
\caption{Origin of the enhancement of 
$M_3(\psi)$ (the $3\omega$ harmonic
of $\cal{M}_\bot$)
at $h=h_{ac}=0.05$, and 
$\psi\equiv 
\psi_0=25^0$ (recall Fig. \protect\ref{mom3}) with the admixture of 
$s$-wave component 
in the $d+is$ gap. The curves, which represent the product
of the normalized magnetic 
moment ${\cal M}_\bot$ and $\cos(3 \omega t)$,  (see Fig. 
\protect\ref{momx})
are labeled by the corresponding value of $\delta$.}
\label{momt}
\end{figure}
\begin{figure}
\caption{Field ratio, $R\equiv h_{dc}/h_{ac}$, 
dependence of $M_{2}^4$
(the angular 
$\pi/2$ Fourier 
component of the $2\omega$ signal of
the  normalized transverse magnetic moment 
${\cal M}_\bot$ , Eq. (\protect\ref{m3ij})), 
for various admixtures of $s$-wave in a 
$d+is$ energy gap. Results are given at 
$h=0.05$. The normalization is as in Fig. \protect\ref{mom3}. 
Curves are 
labeled by the ratio $\delta=\Delta_s/\Delta_d$ with the same values as 
in Fig. \protect\ref{mom3}.}
\label{opt2}
\end{figure}
\begin{figure}
\caption{Field ratio, $R$, dependence of the angular $\pi/2$ Fourier 
component, $M_{3}^4$, of the $3 \omega$ signal of
the normalized transverse magnetic moment, 
${\cal M}_\bot$. Results shown are
 for various admixtures of $s$-wave in a 
$d+is$ state at $h=0.05$. Normalization and labeling are
the same as in Fig. \protect\ref{opt2}.}
\label{opt3}
\end{figure}
\begin{figure}
\caption{Dependence of the angular 
$\pi/2$ Fourier component, $M_{2}^4$, of the $2\omega$
nonlinear magnetic moment signal
${\cal M}_\bot$ (recall Fig. \protect\ref{opt2}) on the 
admixtures of $s$-wave component in a $d+is$ gap. Results are shown at 
$h=0.05$ and normalized as in  
Fig. \protect\ref{opt2}. Curves plotted are labeled by the 
corresponding value of $R$.}
\label{del2}
\end{figure}
\begin{figure}
\caption{Dependence of the angular  $\pi/2$ Fourier 
component of the $3\omega$ signal, 
$M_{3}^4$, of ${\cal M}_\bot$ (compare with  Fig. 
\protect\ref{opt3}) on the 
admixture of $s$-wave component in a $d+is$ state. Results are shown at 
$h=0.05$ and normalized 
as in Fig. \protect\ref{opt2}. 
Curves plotted are labeled by their value of $R$.}
\label{del3}
\end{figure}

\begin{references}
\bibitem{fsa}We assume here a cylindrical (not necessarily isotropic) FS.
This is appropriate for unconventional layered superconductors.
As shown in Ref. \protect{\onlinecite{zv}}, inclusion of a 
non-cylindrical 3D FS introduces only a small correction.
\bibitem{weight} For an  anisotropic FS surface some care has to be taken: 
at the nodal position $s_n$, the density of states, $n(s_n)$, is not 
necessarily equal to the average value, $n_{iso}$, at each node we define 
$N_{fn} \equiv N_f n(s_n)/n_{iso}$.
\bibitem{thick} It is simple to extend the solution to
arbitrary thickness, as shown in Refs. \protect{\onlinecite{sv,ys,zv3}}.
\bibitem{sid} For our purposes there is no distinction between a $d+is$ 
and the $s+id$ gap considered in Ref. \protect\onlinecite{zv3} as we 
discuss results which only depend on the absolute value of the OP.
\bibitem{norm} The normalization factor is the angular $\pi/2$ Fourier 
component of the normalized angular dependence for a pure $d$-wave 
$(1/(3\sqrt{3}) \cos \psi \sin \psi (\cos \psi-\sin \psi))$, very 
approximately equal to unity.
\end{references}
\end{document}